\documentclass[prl,twocolumn]{revtex4}
\usepackage{graphicx}
\renewcommand{\narrowtext}{\begin{multicols}{2} \global\columnwidth20.5pc}
\renewcommand{\v}[1]{{\bf #1}}

\def\be{\begin{eqnarray}}
\def\ee{\end{eqnarray}}
\newcommand{\nn}{\nonumber\\}
\newcommand{\Eq}[1]{Eq.~(\ref{#1})}
\newcommand{\Fig}[1]{Fig.~(\ref{#1})}

\begin{document}

\title{Renormalization Group Study of the Electron-phonon Interaction in High $T_c$ Cuprates}

\author{Henry C. Fu$^{a}$, Carsten Honerkamp$^{b,c}$, and Dung-Hai Lee$^{a,d,e}$}
\affiliation{${(a)}$Department of Physics,University of California
at Berkeley, Berkeley, CA 94720, USA}
\affiliation{${(b)}$Institute for Theoretical Physics, Universit\"at W\"urzburg, D-97074 W\"urzburg, Germany}
\affiliation{${(c)}$Max-Planck-Institute for Solid State Research, D-70569 Stuttgart, Germany}
\affiliation{${(d)}$ Material Science
Division, Lawrence Berkeley National Laboratory,Berkeley, CA 94720,
USA.}
\affiliation{$(e)$ Center for Advanced Study, Tsinghua
University, Beijing 100084, China.}

\date{\today}

\begin{abstract}
We generalize the numerical renormalization group scheme of Ref.
\cite{Carsten,Zanchi,Halboth} to study the phonon-mediated
retarded interactions in the high $T_c$ cuprates. We find that
three sets of phonon-mediated retarded quasiparticle scatterings
grow under RG flow.  These scatterings share the following common
features: 1) the initial and final quasiparticle momenta are in
the antinodal regions, and 2) the scattering amplitudes have a
$x^2-y^2$ symmetry.  All three sets of retarded interaction are
driven to strong coupling by the magnetic fluctuations around
$(\pi,\pi)$. After growing strong, these retarded interaction will
trigger density wave orders with d-wave symmetry. However, due to
the d-wave form factor they will leave the nodal quasiparticle
unaffected. We conclude that the main effect of electron-phonon
coupling in the cuprates is to promote these density wave orders.
\end{abstract}

\maketitle


One of the most prominent puzzles to emerge from experimental
studies of the high-temperature superconducting cuprate materials
is the strong momentum-space dependence of the properties of the
electronic excitations. This can be seen most clearly in
angle-resolved photoemission spectroscopy (ARPES) of LSCO: at
optimum doping, sharp quasiparticle peaks can be distinguished near
the Fermi surface in both nodal and antinodal directions.  However,
when the system is underdoped, the antinodal quasiparticle peaks
disappear, while the nodal quasiparticle peaks
remain\cite{Zhou-lsco}.  
Another beautiful demonstration of the robustness of the nodal
quasiparticle is obtained in the NaCCOC materials. Recent STM
studies have shown the presence of a commensurate, 4 lattice
constant, checkerboard order which is independent of
doping\cite{Hanaguri} (indicating the importance of lattice
pinning). Despite this, ARPES studies of the same system
reveal nodal quasiparticle peaks\cite{KShen}. %
Finally, ARPES experiments probing the isotope effect in optimally
doped Bi2212 reveal a substantial isotope shift in the ARPES spectra
near the antinode, but little shift in the nodal
spectrum\cite{Gweon}. The broad picture emerging from these
experiments is of nodal quasiparticles which are insensitive to
doping, disorder, charge and spin order, and lattice vibration,
while antinodal excitations are sensitive to all these
perturbations. One of us (DHL) has dubbed this the ``nodal-antinodal
dichotomy''\cite{Zhou-lsco}.

In the cuprates it is widely accepted that in the overdoped
regime, quasiparticles are well-defined excitations all along the
normal state Fermi surface.  We view the effect of decreasing
doping as changing the Fermi surface geometry and increasing the
strength of the residual quasiparticle interaction. We model this
residual interaction with a momentum independent
quasiparticle scattering. 
It is important not to confuse this effective residual
quasiparticle scattering with the bare local electron correlation.
Previously Honerkamp {\it et al} have performed a
one-loop renormalization group (RG) study of this residual
interaction\cite{Carsten}. Here we generalize their method to
study the phonon-mediated retarded interaction. The RG treatment
of the retarded interaction is similar to that used for one
dimensional systems in Ref.\cite{Zimanyi,Seidel,3leg}. Our
motivation for studying the electron-phonon interaction in the
cuprates are twofold: 1) we hope to better understand the
electron-phonon interaction in doped Mott insulators, and 2) we
hope to gain some insight about the origin of the nodal-antinodal
dichotomy.

In a recent paper Devereaux {\it et al} pointed out the importance
of the momentum dependence in the electron-phonon coupling constant
when interpreting ARPES data\cite{Devereaux}. In particular they
showed that as the initial electron momentum is varied along the
Fermi surface, the matrix element that couples the electron to the
B$_{1g}$ phonon exhibits a $x^2-y^2$ symmetry. Meanwhile studies
including weak to intermediate Hubbard interaction ($U<6t$) 
have come to the
conclusion that the s-symmetry electron-phonon coupling is
suppressed by electron-electron
repulsion\cite{Kim,Grilli,Zeyher,Huang}, especially for large
momentum transfer processes\cite{Grilli,Zeyher,Huang}.

\begin{figure}
\includegraphics[width=7.5cm]{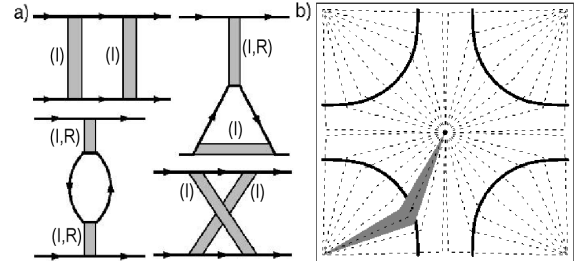}
\caption{a) One loop Feynman diagrams contributing to
$\partial_\Lambda V_{\Lambda}$. Spin is conserved along solid lines.
The labels ``I'' and ``R''  represent the instantaneous and retarded
interaction (the gray bars) respectively. While all interactions can
be the ``I'' type,
only those with both ``I'' and ``R'' labels can be retarded.
A diagram contributes to the flow of the instantaneous interaction
when all interactions are  the ``I'' type; otherwise it contributes to the flow of the retarded interaction.
b) The 32-patch discretization of the Fermi surface. Each patch
(example in gray) is centered around one of the dashed
lines.}\label{diagrams}
\end{figure}

In this study we follow the RG flow of both the instantaneous and
retarded electron-electron scattering amplitude in the full
first Brillouin zone.  We compare the flow of retarded
interaction with different symmetry. We start with a tight binding
dispersion given by $ \epsilon({\bf k}) = -2t[\cos(k_x) + \cos(k_y)]
+ 4 t' \cos(k_x) \cos(k_y) + 4 t'' [\cos^2(k_x) + \cos^2(k_y) -1],$
with $t' = 0.3 t$, $t'' = -0.1 t$, and $\mu = -0.7 t$.  This set of parameter choice produces a rather realistic Fermi surface.  Following Ref.\cite{Carsten} we choose the bare residual quasiparticle scattering amplitude $U=3t$.  As in Ref.\cite{Carsten} we perform a one-loop Wilsonian RG for the
one-particle-irreducible four-point vertex functions. At each energy
scale the renormalized four-point function serves as an effective
interaction for particles with excitation energies $|\epsilon(\bf
k)|$ below the cutoff scale $\Lambda$. The results we report
are obtained for a temperature $k_BT=0.04t$, and the RG flow is
integrated between initial cutoff $\Lambda=4t$ and final cutoff $0.2
t$. With spin-rotational invariance, the cutoff-dependent effective
interaction is given by \be S_{int} = &&\sum_{\sigma,\sigma'}\int
{\prod_{i=1}^3} d^3 {\bf k_i} d^3 \omega_i V_\Lambda ({\bf k_1},{\bf
k_2},{\bf k_3}) \psi_\sigma^{\dagger}({\bf k_4}, \omega_4)\nn&&
\psi_{\sigma'}^{\dagger}({\bf k_3},\omega_3) \psi_{\sigma'}({\bf
k_1}, \omega_1) \psi_\sigma({\bf k_2}, \omega_2).\ee In the above
$\psi_\sigma(\omega_i,\v k_i)$ annihilates an electron with quantum
numbers $\omega_i, \v k_i,\sigma$; furthermore $\omega_4$ = $\omega_1 +
\omega_2 - \omega_3$ and ${\bf k_4} = {\bf k_1}+{\bf k_2}-{\bf
k_3}$.  The contribution to the RG flow $\partial_\Lambda
V_{\Lambda}$ is summarized by the Feynman diagrams in
\Fig{diagrams}, where the label ``I'' or ``R'' denotes instantaneous
or retarded interaction respectively. For the moment we discuss the instantaneous interaction, hence all interactions are
the ``I'' type. In each diagram, there are two internal lines. One
represents the Greens function
\begin{equation}
G_{\Lambda}({\bf k},\omega) = \frac{\chi_{\Lambda}({\bf k})}{i\omega
- \epsilon({\bf k}) - \chi_{\Lambda}({\bf k}) \Sigma({\bf
k},i\omega)}, \label{off}
\end{equation}
while the other represents
\begin{equation}
S_{\Lambda}({\bf k},\omega) = \frac{\chi'_{\Lambda}({\bf k})[i\omega
- \epsilon({\bf k})]}{[i\omega - \epsilon({\bf k}) -
\chi_{\Lambda}({\bf k}) \Sigma({\bf k},i\omega)]^2}.
\label{on}\end{equation} Here $\chi_{\Lambda}(\v k) =1 -1/\{\exp[(
|\epsilon(\v k) |- \Lambda)/0.05 \Lambda] +1\}$ cuts off 
contributions from $|\epsilon(\v k)|<\Lambda$, and $\chi'_\Lambda(\v
k)\equiv
\partial_{|\Lambda|}\chi$ is nonzero only for $|\epsilon(\v k)|$
near the cutoff $\Lambda$. Each diagram stands for two
contributions since there are two ways to assign $G_\Lambda$ and $S_\Lambda$ to the internal lines. 
As in 
previous works\cite{Carsten,Zanchi,Halboth} our calculation
ignores higher order vertices and self-energy corrections (i.e $\Sigma_{\Lambda}$ is set to zero).  Therefore the RG flow has to be stopped before the interactions get too large and the lowest energy scales cannot be accessed in a controlled way.

We perform the RG numerically by dividing the Brillouin zone into 32
patches.  Each patch is centered on a ``radial" line, as shown in
\Fig{diagrams}.  By standard Taylor expansion and power counting
arguments, the most relevant vertex functions are the ones with
incoming and outgoing $\v k_i$ on the fermi surface and the
$\omega_i=0$. In the following we only keep track of these most
relevant vertex functions, which means we ignore the momentum
dependence in directions transverse to the Fermi surface. For the
instantaneous electron-electron scattering we thus approximate the
value of $V_{\Lambda}({\bf k_1}, {\bf k_2}, {\bf k_3})$ by its value
$V_{\Lambda}(\v k_F(i), \v k_F(j), \v k_F(l))$, where $\v k_F(j)$ is
the momentum on the fermi surface at
the center of the jth patch. 
This leaves us with $32^3$ couplings $V(\v k_F(i), \v k_F(j),\v
k_F(l))$ to keep track of.
We calculate the contributions to the flow of these couplings using
the diagrams in \Fig{diagrams}. The frequency sums are performed
analytically, and the loop momentum integration is performed by a
sum over patches and integration along the radial lines.

The retarded interactions have non-trivial frequency dependence. We
assume them to have the form
\begin{equation}
V_R({\bf k_1}, \omega_1; {\bf k_2}, \omega_2; {\bf k_3}, \omega_3) =
-\frac{g({\bf k_1}, {\bf k_3}) g({\bf k_2},{\bf k_4})
\Omega_D}{(\omega_1 - \omega_3)^2 + \Omega_D^2}.\label{reint}
\end{equation}
Here $g({\bf k_1},{\bf k_3})$ is the electron-phonon matrix element
for scattering an electron from momentum ${\bf k_1}$ to ${\bf k_3}$,
and $\Omega_D$ is a characteristic phonon (Debye) frequency. The
retarded interaction dies off for frequency transfer $|
\omega_1-\omega_3 | > \Omega_D$.  In this work we approximate this
dependence by a step-function cutoff\cite{Zimanyi}.  This approximation throws away the full-frequency dependence of the retarded electron interaction used in other works\cite{Tsai} but makes the study of experimentally relevant fermi surface shapes calculationally tractable.  
This retarded interaction
is represented by a new set of $32^3$ ``retarded" couplings,
$V_R(\v k_F(i),\v k_F(j),\v k_F(l))$.
Because the internal lines in
each of the diagrams have frequency poles at or greater than the cutoff energy $\Lambda$ while those of the external legs are at zero energy,
in the limit $\Lambda >> \Omega_D$ the leading contributions to the RG flow of the retarded interaction comes from diagrams in \Fig{diagrams} where at least one
interaction line is the  ``R" type.    The contributions from diagrams with retarded interactions placed at vertices without an ``R" label in \Fig{diagrams} are down by a factor
$\Omega_D/\Lambda$.  Because the phonon energies of interest in
the cuprates are $\sim 60$ meV\cite{Cuk}, in this work we
integrate the RG flow equations to a lower cutoff of $0.2 t$.
Since the retarded interaction is present only when the frequency
transfer is smaller than $\Omega_D$, diagrams with retarded
interactions contribute only to the flow of retarded vertex
function. 
Consequently the flow of instantaneous couplings is unaffected by
the addition of retarded couplings, but the flow of retarded
couplings is affected by the instantaneous interactions.

\begin{figure}
\includegraphics[width=7.5cm]{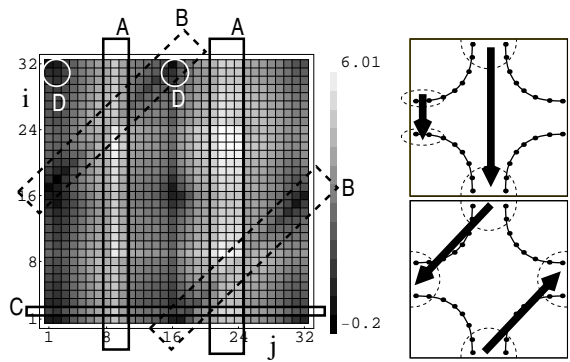}
\caption{The renormalized instantaneous interaction
$V_{\Lambda=0.2t}(\v k_F(i), \v k_F(j), \v k_F(l))$ plotted as a
function of $i$ and $j$ for a fixed $l=2$.  The solid vertical boxes 
(A) enclose
couplings which promote commensurate and incommensurate magnetic
ordering around $(\pi,\pi)$. Examples of these scattering processes
are shown on the right with each dashed arc enclosing a segment of the
Fermi surface involved in the scattering processes. The dashed diagonal boxes (B) enclose couplings which promote d-wave superconductivity. 
The solid horizontal box (C) encloses couplings which lead to the Pomeranchuk instability.  The circles (D) enclose couplings which lead to 4 lattice constant charge-density-wave order.   \label{elecresults}}
\end{figure}

The RG flow of the instantaneous quasiparticle
scattering alone with band parameter $t'' =0$ has been obtained by
Honerkamp et. al. We obtain similar results  when including $t''=
-0.1t$.  Depending upon the doping level, several groups have
found that the renormalization flow enhances instantaneous
couplings which favor either SDW formation or d-wave
superconductivity as the most prominent ordering
tendency\cite{Carsten,Halboth,Zanchi}. These can be seen in
\Fig{elecresults} where $V_{\Lambda=0.2t}(\v k_F(i), \v k_F(j), \v
k_F(l))$ is plotted as a function of $i$ and $j$ for a fixed
$l=2$.
The couplings which favor SDW ordering have a positive amplitude and
constant momentum transfer $\v k_F(j)-\v k_F(l)$\cite{Carsten}. They
show up as the vertical bands in boxes (A) of
\Fig{elecresults}. The values $\v k_F(j)-\v k_F(l)$ of these
scattering processes are around $(\pi,\pi)$. In particular they
include $(\pi \pm 2\pi/8 a, \pi)$, indicating ``incommensurate" SDW
ordering. The couplings which favor d-wave superconductivity are the
diagonal lines enclosed in boxes (B) of
\Fig{elecresults}. These scatterings occur in the cooper channel
where $\v k_F(i) + \v k_F(j)= 0$, and have an alternating sign
structure indicative of d-wave symmetry. The fact that the couplings
favoring superconducting pairing overlap with the couplings favoring
SDW ordering signifies that physically, these ordering tendencies
are linked.
{\it Since the phonon-mediated retarded couplings do not influence the flow of the instantaneous interactions, the d-wave superconducting pairing seen
above can not be due to the electron-phonon interaction.} In the
present calculation, due to the choice of chemical potential and $U$, the spin-density waves are always the leading ordering tendency.

In addition to these two types of interactions, there also exist
weaker scattering processes which favor other types of
ordering. Couplings which lead to these ordering tendencies are also indicated in \Fig{elecresults}. The first is the Pomeranchuk instability, which leads to a
$x^2 - y^2$-symmetry deformation of the Fermi surface\cite{Halboth}.
 The second
leads to a charge-density wave order with ordering wavevector equal
to the vectors connecting the parallel segments of the Fermi surface
near the Brillouin zone face.  These wavevectors correspond
to a spatial period near four lattice constants.  It is worth
noting that the $t'$ and $t''$ terms in the band structure are
responsible for the almost nested Fermi surface near the Brillouin
zone face.  They enhance the charge ordering tendency.

\begin{figure}
\includegraphics[width=7cm]{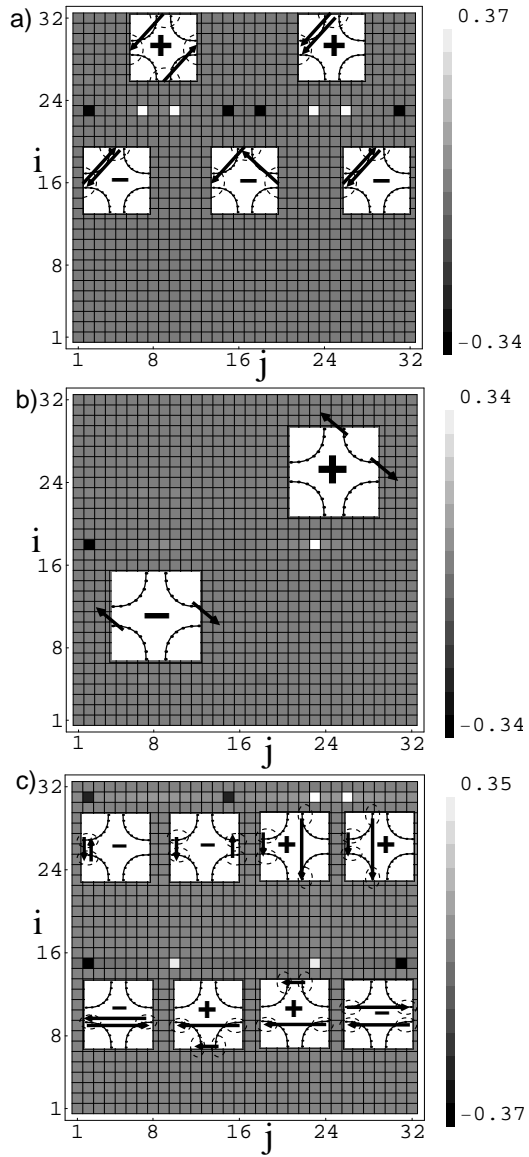}
\caption{ The three enhanced sets of retarded interactions, with
their major types of scattering processes plotted as in
\Fig{elecresults}.  $\v k_F(l)$ is fixed at $l=2$.  Each set of
retarded interactions has both positive and negative couplings.  The
sign structure is indicative of a $x^2-y^2$ symmetry in momentum
space.} \label{retinstability}
\end{figure}

Now we investigate the flow of the retarded couplings.
By experimenting with the symmetries of the initial scattering
amplitude, we find three independent sets of retarded interactions
that are most
strongly enhanced by the RG (\Fig{retinstability}). 
In all three sets the scattering amplitude shows $x^2-y^2$
symmetry in momentum space. This type of sign structure can arise
from phonon mediation if 
$g({\bf k}, {\bf k+Q})$ in \Eq{reint} transforms as
$cos(k_x)-cos(k_y)$.  For the interaction in Figs. (3a), (3b), and (3c), ${\bf Q} \sim (\pm \pi, \pm \pi)$, ${\bf Q} \sim ( \pm 2 \pi/4a,\pm 2
\pi/4a)$, and ${\bf Q} \sim (\pm 2 \pi/4a,0),(0,\pm 2 \pi/4a)$,
respectively. For example, it can arise from coupling to the half breathing or the B$_{1g}$ \cite{Devereaux} phonons.  Note that the electron-phonon interaction with the B$_1g$ buckling mode has no amplitude for momentum transfer $(\pi,\pi)$\cite{Jepsen} and so will not contribute appreciably to retarded interaction (3a).  Consistent with previous
results\cite{Kim,Grilli,Zeyher,Huang}, the same analysis indicates
that an s-symmetry electron-phonon coupling will be suppressed
under RG. 
(Monte Carlo studies\cite{Huang} find that s-symmetry electron-phonon coupling grows in the Hubbard model for large $U>6t$; it would be of interest to see the behavior of $x^2 -y^2$ couplings.)      
Another common feature among the three sets is that the involved initial and final momenta are all within the antinodal region.

We find that the most important contribution to the RG flow of the
retarded interaction comes from the ``bubble" diagram in the lower left of
\Fig{diagrams}(a), when one of vertices is an instantaneous interaction from the boxes (A) of \Fig{elecresults} and the other is a retarded interaction. The $x^2-y^2$ sign structure discussed above
is essential for the growth of the scattering amplitude: the
positive(negative) retarded coupling combines with the instantaneous
interactions to drive the negative(positive) retarded interaction to
strong coupling.  Since the boxes (A) enclose interactions that
favor SDW ordering near $(\pi,\pi)$, we conclude that {\it
commensurate and incommensurate magnetic fluctuations are
responsible for driving the relevant retarded interactions to strong
coupling.}

The key features of each set of retarded interactions in
\Fig{retinstability} appear as horizontal bands.  In a
horizontal band the momentum transfer $\v k_F(i)-\v k_F(l)$ is
fixed, suggesting that these couplings promote some type of
density wave order. However, the $x^2-y^2$ sign structure
makes the corresponding density wave order an unconventional one.
The density wave order parameter driven by the retarded
interactions in \Fig{retinstability} has the form 
\begin{equation} {\cal O} =  \sum_{\v k,\sigma}
f(\v k) < c^{\dagger}_{{\bf k+Q}\sigma} c_{\bf k\sigma}>,
\label{dCDW}
\end{equation}
where $f(\v k)$ transforms like $\cos(k_x)-\cos(k_y)$ under
rotation.  These are examples of generalized d-wave density
orders, which in general can have bond averages $<c^{\dagger}_i
c_j>$ with any complex phase, leading to both bond current and
charge density orders\cite{Chakravarty}.  We have verified via mean field
calculations (details in a later publication) that all three 
interactions can promote real space modulations in both charge density 
and current.
In particular, the interaction of Fig.(3c) can lead to periodic
charge density modulations with period $\sim 4a$.  Note that the
chemical potential $\mu$ can change the nesting wavevector and
hence the period of modulations.
Thus we find that in the cuprates {\it the most important effect
of the electron-phonon interaction is to promote density wave
order.} Interestingly, for any order described by \Eq{dCDW}, the
nodal quasiparticles will be unaffected. In particular the density
wave order in Fig.(3c) is consistent with the STM experiment of
Hanaguri {\it et al} and the ARPES experiment by Shen {\it et al}.
In addition, the fact that all the enhanced retarded interactions
in \Fig{retinstability} involve scattering processes with initial
and final momentum states in the antinodal region is also
consistent with the isotope-dependent ARPES study of
Ref.\cite{Gweon}.


Acknowledgement: We are in debt to J.C. Davis, K.P. McElroy, G.-H.
Gweon, A. Lanzara, K. Shen, Z.-X. Shen, D. Scalapino, W. Hanke, and W. Metzner  for many discussions. DHL is supported by DOE grant DE-AC03-76SF00098.  HCF was partially supported
by a predoctoral fellowship from the Advanced Light Source.


\end{document}